\documentclass[12pt]{spieman}  
\usepackage{amsmath,amsfonts,amssymb}
\usepackage{graphicx}
\usepackage{setspace}
\usepackage{tocloft}
\usepackage{lineno}
\usepackage{doi}
\usepackage{siunitx}
\usepackage{booktabs}
\usepackage{multirow, array}
\usepackage{lineno}
\usepackage{subcaption}
\usepackage{float} 
\usepackage{xcolor}
\usepackage{soul}
\usepackage{tablefootnote}

\newcommand{\apj}{ApJ}
\newcommand{\apjl}{ApJL}

\newcommand{\solphys}{Solar Physics}

\DeclareSIUnit\keV{\kilo\electronvolt}
\DeclareSIUnit{\counts}{counts}
\DeclareSIUnit{\channel}{channel}
\DeclareSIUnit\erg{erg}

\title{Solar Low Energy X-ray Spectrometer on board Aditya-L1: Ground Calibration and In-flight Performance}

\author[a,b,*]{Abhilash R. Sarwade}
\author[a]{Ankur Kushwaha}
\author[a]{M. C. Ramadevi}
\author[a]{Monoj Bug}
\author[a]{Kiran Lakshmipathaiah}
\author[a]{Smrati Verma}
\author[a]{Vaishali Sharan}
\author[a,c]{K. Sankarasubramanian}
\affil[a]{U. R. Rao Satellite Centre, ISRO, Bengaluru, Karnataka, India}
\affil[b]{Indian Institute of Technology Guwahati, Guwahati - 781039, Assam, India}
\affil[c]{Centre for Excellence in Space Science and Instrumentation, IISER-Kolkatta, Kolkatta,
India}

\cftpagenumbersoff{figure}
\cftpagenumbersoff{table} 
\begin{document} 
\maketitle

\begin{abstract}
The Solar Low-Energy X-ray Spectrometer (SoLEXS) on board India's Aditya-L1 mission was launched on 2 September 2023 and commenced solar observations on 13 December 2023 following successful aperture cover deployment. Operating from the Sun-Earth L1 Lagrange point, SoLEXS has been providing continuous Sun-as-a-star soft X-ray spectroscopy across \SIrange{2}{22}{\keV} with \SI{170}{\electronvolt} resolution at \SI{5.9}{\keV} and \num{1}-second temporal cadence since 6 January 2024. The instrument employs two Silicon Drift Detectors with aperture areas of \SI{7.1}{\milli\metre\squared} and \SI{0.1}{\milli\metre\squared} to accommodate the full dynamic range of solar activity from A-class to X-class flares.
	
This paper presents comprehensive ground and on board calibration procedures that establish SoLEXS's quantitative spectroscopic capabilities. Ground calibration encompassed energy-channel relationships, spectral resolution characterization, instrument response functions, and collimator angular response measurements, with thermo-vacuum testing validating performance stability across operational temperature ranges. On board calibration utilizing an internal $^{55}$Fe source demonstrated preserved post-launch spectral resolution (\SIrange{164.9}{171.2}{\electronvolt}), while cross-calibration with GOES-XRS and Chandrayaan-2/XSM confirmed radiometric accuracy and flux agreement.

The instrument's \SI{100}{\percent} observational duty cycle at L1 enables unprecedented continuous monitoring of solar flare evolution across all intensity classes, providing calibrated data for advancing coronal heating mechanisms, flare energetics, and flare-coronal mass ejection relationship studies through soft X-ray spectroscopy.
\end{abstract}


{\noindent \footnotesize\textbf{*}Abhilash R. Sarwade,  \linkable{sarwade@ursc.gov.in} }

\begin{spacing}{1}   

\section{Introduction}
\label{sect:intro}  
The solar corona, characterized by million-degree plasma ($T \sim $\SIrange{1}{3}{\mega\kelvin}), far exceeds the temperatures of the underlying photosphere and chromosphere. 
During explosive solar flares, coronal temperatures surge to $T \sim $\SIrange{10}{20}{\mega\kelvin} as magnetic reconnections release energy up to \SI{1e32}{\erg}\cite{Shibata2011}. 
Solar soft X-ray spectroscopic observations have been used for decades to probe coronal plasma parameters such as temperature, density, and elemental abundances through the diagnostic line and continuum emissions \cite{delzannaSolarUVXray2018}.
These spectroscopic observations are crucial to understanding solar coronal heating mechanisms and flare energetics \cite{Parker1988,Klimchuk2006,Aschwanden2015}. The spectroscopic capability started with pioneering rocket-borne experiments \cite{Friedman1951,blake1964solar} and early orbital platforms like the Orbiting Solar Observatory series (1962-1975) \cite{Walker1972,Neupert1971}, which established the foundation for solar X-ray studies.

Fundamental trade-offs between spectral resolution, energy coverage, and observational duty cycles have shaped the evolution of solar X-ray spectroscopy. Bragg crystal spectrometers such as Solar Flare X-rays (SOLFLEX) on P78-1 (1979-1981) \cite{Doscheck1981}, Solar Maximum Mission (SMM) (1980, 1984-1989) \cite{ActonSMMXRP1980} and Yohkoh (1991-2001) \cite{CulhaneYohkohBCS1991}, RESIK (REntgenovsky Spectrometer s Izognutymi Kristalami) on CORONAS-F (2001-2005) \cite{Sylwester2005} achieved high spectral resolution ($\Delta\lambda/\lambda \sim$ \num{1e-3} or $\Delta$E $\sim$ \SI{10}{\electronvolt}) through mechanical scanning of narrow wavelength bands. However, these instruments targeted specific emission lines with limited energy range \cite{Phillips2004b}.

The Reuven Ramaty High Energy Solar Spectroscopic Imager (RHESSI) (2002-2018) \cite{Lin2002} expanded observations into hard X-rays (\SI{3}{\keV}--\SI{17}{\mega\electronvolt}) using Germanium detectors, providing insights into flare energetics by studying thermal and nonthermal emission \cite{SaintHilaire2005,Aschwanden2016,Warmuth2016}. However, its coarse spectral resolution of $\sim$ \SI{1}{\keV} at \SI{6}{\keV} limited detailed abundance analysis \cite{Phillips2012} and its sensitivity to quiescent soft X-ray emissions \cite{McTiernan2009}.

The next generation of instruments, Solar X-ray Spectrometer (SOXS) on GSAT-2 (2004-2011) \cite{Jain2005} and Solar Photometer in X-rays (SphinX) on CORONAS-Photon (2009) \cite{Gburek2012}, employed Silicon PIN diodes to achieve improved energy resolutions (\num{0.7} and \SI{0.4}{\keV} at \SI{6}{\keV} for SOXS and SphinX respectively), lower energy detection thresholds (\SI{1.2}{\keV} for SphinX), and enhanced temporal resolution (sub-seconds). These capabilities enabled more detailed abundance analysis and studies of temporal evolution during flares \cite{Jain2006,Mrozek2012}.

Modern Silicon Drift Detectors (SDDs) represent a new type of detector that achieves better spectral resolution ($\sim$ \SI{170}{\electronvolt} at \SI{5.9}{\keV}) with broadband energy coverage of \SIrange{1}{30}{\keV} and higher count rate capabilities than Si PIN detectors. The SDD architecture features low anode capacitance independent of detector area, integrated on-chip electronics for noise minimization, and pulsed reset capability enabling high count rate operation without spectral degradation \cite{Niculae2006}.

Recent SDD-based instruments like Miniature X-ray Solar Spectrometer (MinXSS) (2015-2017; 2018-2019) \cite{MooreMINXSS2018}, Dual Aperture X-ray Solar Spectrometer (DAXSS) (2022-2023) \cite{Woods2023}, Solar X-ray Monitor (XSM) on Chandrayaan-2 (2019-present) \cite{Mithun2020_xsm}, and Solar X-ray Detector (SXD) on Macau Science Satellite-1B (MSS-1B) (2023-present) \cite{Ng2024} demonstrated enhanced SDD capabilities.
For example, XSM used SDD in the study of quiet-Sun \cite{Vadawale2021QuietSun}, microflares \cite{Vadawale2021microflares}, elemental abundance evolution \cite{Mondal2021}, multithermal plasma analysis \cite{mithunSoftXRaySpectral2022}, while MSS1B's SXD captured the evolution of low FIP abundances during X-class flares \cite{Ng2025}.

Previous solar X-ray spectrometers' critical limitation is their orbital duty cycles, which significantly impact observational continuity and data quality. Instruments in low Earth orbit face regular eclipses and South Atlantic Anomaly transits, cumulatively reducing effective observing time to \SI{<70}{\percent} \cite{Inglis2011}. 
In contrast, the Sun-Earth L1 Lagrange point provides continuous solar visibility, a vantage point utilized by solar missions like SOHO, ACE, and WIND \cite{Roberts2011}, though none previously incorporated SDD-based soft X-ray spectrometers.

The Solar Low-Energy X-ray Spectrometer (SoLEXS) on board Aditya-L1 \cite{Sankarasubramanian2025} combines the advantages of SDD technology with continuous solar viewing at the Sun-Earth L1 Lagrange point. The instrument features dual SDDs with \SI{7.1}{\milli\metre\squared} and \SI{0.1}{\milli\metre\squared} aperture areas covering the \SIrange{2}{22}{\keV} range.
This configuration captures the full dynamic range of solar activity from A-class to X-class flares with \SI{1}{\second} temporal resolution and energy resolution $\approx$ \SI{170}{\electronvolt} at \SI{5.9}{\keV}. SoLEXS's scientific objectives are: 
(i) quantifying the contribution of ubiquitous small-scale flare events (e.g., microflares, nanoflares) to steady coronal heating,
(ii) correlating flare plasma properties with coronal mass ejections (CMEs), and 
(iii) measuring temperature, emission measure, and abundance evolution via flare spectroscopy.

Achieving these science objectives requires instrument calibration spanning pre-launch characterization and in-flight validation. The calibration process ensures that the measured spectra can be reliably interpreted in terms of incident solar X-ray fluxes, energy spectrum, and temporal variability. Ground calibration activities encompass spectral response characterization and assessment of detector performance across the full range of operational temperatures and count rates. On board calibration utilizes an internal radioactive $^{55}$Fe source for gain and spectral resolution stability monitoring, complemented by cross-calibration with concurrent XSM and X-Ray Sensor (XRS) on Geostationary Operational Environmental Satellites (GOES) \cite{Chamberlin2009} observations.

This paper presents the ground and in-flight calibration of the SoLEXS instrument, covering its design, digital pulse processing, and ground-based characterization of its instrument response as well as thermo-vacuum performance. 
It further presents the on board performance verification, cross-calibration with GOES/XRS and Chandrayaan-2/XSM, and a demonstration of the derivation of coronal plasma parameters from flare spectra.

\section{SoLEXS Instrument} \label{sect:instrument}

Solar Low Energy X-ray Spectrometer (SoLEXS) \cite{Sankarasubramanian2025} is one of the payloads on board Aditya-L1, capable of performing full-disk (Sun-as-a-star) spectrometry in the energy range of \SIrange{2}{22}{\keV}.
The instrument is configured employing two SDDs, aligned with respective apertures of optimized diameters.
This two aperture configuration ensures that at least one detector operates below saturation flux level despite a probable \num{\sim 5} orders of magnitude in solar X-ray flux variation from faint A-class flares to bright X-class events during mission life.

\begin{figure}
	    \centering
	    \includegraphics[width=1\linewidth]{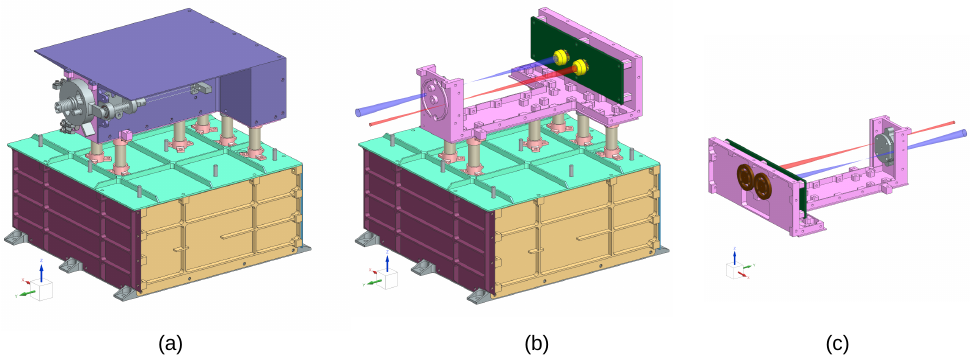}
	    \caption{
        Schematic of SoLEXS instrument, 
        (a) A front view of the integrated SoLEXS package, showing the Detector Module (top) and the Electronics Module (bottom), which are thermally isolated by GFRP spacers. Aperture cover mechanism (grey) is visible on the outside of the detector module. 
        (b) An internal front view of the package with the detector module's top cover removed, revealing the two SDDs (yellow) mounted on the PCB (green). The imaginary FOVs are shown originating from the two apertures on the front plate: the blue FOV for the large-aperture SDD1 and the red FOV for the small-aperture SDD2.
        (c) A close-up back view of the detector module with the top cover removed, showing the internal calibration source encapsulated in its Aluminium holder (grey) and the Copper thermal flanges (brown). 
        }
	    \label{fig:solexs_front}
	\end{figure}

The mechanical design of SoLEXS consists of two modules; a detector module, with two SDDs as well as a calibration source, and an electronics module for signal processing, power management, and spacecraft interfacing. These two modules are thermally isolated by seven glass-fibre-reinforced polymer (GFRP) spacers but maintain mechanical integrity (see Figure~\ref{fig:solexs_front}).

The detector module consists of a bottom U-shaped cover and the top cover, both fabricated with Al6061-T65 alloy. Two identical SDDs are mounted on a single printed circuit board (PCB), which is fixed on bottom cover as shown in Figure~\ref{fig:solexs_front}b. On the outside of the detector module, a shape memory alloy (SMA) based mechanism covers the apertures during the Earth-bound and cruise phases to minimize the direct radiation damage to the detectors (Figure~\ref{fig:solexs_front}a).
    
The detectors operate at approximately \SI{-45}{\celsius}, achieved through integrated two-stage Peltier coolers within each individual SDD. Copper thermal flanges are mechanically fastened to the detector studs through the back plate (Figure~\ref{fig:solexs_front}c). The heat is extracted from the detectors through the copper flanges to the Aluminium top cover. The top cover, with its extended Aluminium surface area, radiates the heat into deep space, minimizing the active cooling power consumption.

The field of views (FOVs) of SDDs are restricted by two apertures on the opposite side of the bottom U cover (see Figure~\ref{fig:solexs_front}b). The apertures are realized by machining two cylindrical holes of diameters \SI{3.008\pm 0.001}{\milli\metre} and \SI{0.368\pm 0.001}{\milli\metre} (areas of \SI{7.1063 \pm 0.0095}{\square\milli\metre} and \SI{0.1065 \pm 0.0006}{\square\milli\metre}) adjacent to the calibration source. The detectors aligned with these apertures are referred to as SDD1 and SDD2, respectively. Detailed description of the instrument is available in Ref.~\citenum{Sankarasubramanian2025}.

An on board $^{55}$Fe radioactive calibration source, manufactured at the Radiopharmaceuticals Division of Bhabha Atomic Research Centre (BARC) (\url{https://www.barc.gov.in}), is mounted inside the detector module, which irradiates the SDDs through a \SI{2}{\milli\metre} aperture in its Aluminium  holder positioned \SI{135}{\milli\metre} from the detectors (Figure~\ref{fig:solexs_front}c). $^{55}$Fe decays to $^{55}$Mn via electron capture, producing characteristic Mn K$\alpha$\,(\SI{5.898}{\keV}) and Mn K$\beta$\,(\SI{6.490}{\keV}) X-rays. 
Additionally a \SI{6}{\micro\metre} Titanium (Ti) foil is positioned in the beam path that generates fluorescent Ti K$\alpha$\,(\SI{4.507}{\keV}) and Ti K$\beta$\,(\SI{4.932}{\keV}) lines. This calibration source is used to monitor detector performance and energy calibration throughout the mission.

The electronics module contains a stack of four PCBs for low-noise charge-sensitive pre-amplification, front-end and processing electronics, power distribution, and power generation. Initial signal amplification occurs within the detector, with further amplification by charge-sensitive pre-amplifiers in the electronics module. Analog signals are digitized by an ADC and processed by a field-programmable gate array (FPGA), which implements digital pulse processing (DPP) to extract energy and timing information for each incident X-ray photon. Detailed description of the electronics is available in Ref.~\citenum{Bug2025}. The achieved specification of the instrument is given in Table~\ref{table:specification}.

The raw data are transmitted to the Indian Space Science Data Centre (ISSDC) (\url{https://www.issdc.gov.in/adityal1.html}). Higher-level science data products are subsequently generated at the SoLEXS Payload Operations Centre (POC) at U R Rao Satellite Centre (URSC), ISRO (\url{https://www.ursc.gov.in}), which are archived and disseminated via the Pradan web portal (\url{https://pradan.issdc.gov.in/al1}). 
\renewcommand{\thefootnote}{\fnsymbol{footnote}}
\begin{table}[ht]
\centering
\begin{tabular}{ll}
\toprule
\textbf{Parameter} & \textbf{Specification} \\
\midrule
Energy Range & \SIrange{2}{22}{\keV} \\
Energy Resolution & $\approx \SI{170}{\eV}$ at \SI{5.9}{\keV} \\
\multirow[t]{2}{*}[-2pt]{Time Cadence} & Spectral Channel: \SI{1}{\second}\\
 & Temporal Channel: \SI{0.1}{\second} \\
\addlinespace
\textbf{Detector} & \\
\midrule
Type & Silicon Drift Detector (SDD$^{\text{plus}}$)\footnotemark[1] \\
Number & 2 (named SDD1 \& SDD2) \\
Manufacturer & PNDetector \\
Active Area & \SI{30}{\milli\metre\squared} \\
Thickness & \SI{450 \pm 20}{\micro\metre} \\
Entrance Window & \SI{8}{\micro\metre} thick DuraBeryllium Plus\footnotemark[2] \\
Operating Temperature & \SI{-60}{\celsius} to \SI{-40}{\celsius} \\
Internal Collimator Material & Zirconium \\
\multirow[t]{2}{*}{Aperture Area} & SDD1: \SI{7.1063 \pm 0.0095}{\square\milli\metre} \\
                                  & SDD2: \SI{0.1065 \pm 0.0006}{\square\milli\metre}\\
\multirow[t]{2}{*}[-2pt]{Field of View} & SDD1: \SI{\pm 1.8}{\degree}\\
 & SDD2: \SI{\pm 1.3}{\degree} \\
Calibration Source & $^{55}$Fe with Ti foil \\
\addlinespace
\textbf{Digital Pulse Processing Parameters} & \\
\midrule
Pulse Shaping Time (Spectral Channel) & \SI{4}{\micro\second} \\
Pulse Shaping Time (Temporal Channel) & \SI{0.7}{\micro\second} \\
Number of Channels in the Spectrum & \num{340} \\
\multirow[t]{2}{*}[-2pt]{Channel Width} & 1--168 channel: $\approx$\SI{47.75}{\eV}\\
 & 169--340 channel: $\approx$\SI{94.5}{\eV} \\                                                                                                                                                                                                                                                                                     
\addlinespace                                                                        \textbf{Payload Parameters} & \\
\midrule
Mass & $\approx$\SI{3.6}{\kilo\gram} \\
Raw Power & $\approx$\SI{17.9}{\watt} \\
Raw Data Volume & \SI{1.03}{\giga\bit} per 24 hours \\
\bottomrule
\end{tabular}
\caption{Key parameters and specifications of the SoLEXS instrument.}
\label{table:specification}
\end{table}
\footnotetext[1]{\url{https://web.archive.org/web/20250912150300/https://pndetector.de/fileadmin/user_upload/images/brochures/sdd_brochure_2018.pdf}}
\footnotetext[2]{\url{https://web.archive.org/web/20250912150458/https://moxtek.com/wp-content/uploads/pdfs/WIN-DATA-1003-DuraBeryllium-X-ray-Windows-1.pdf}}

\section{Digital Pulse Processing} \label{sect:dpp}
X-ray photons incident on the SDD generate electron-hole pairs proportional to the deposited photon energy within a fully depleted, high-resistivity Silicon volume. The generated electrons are driven toward a small collecting anode by an electric field with a strong component parallel to the detector surface. 
SDD's unique design maintains anode capacitance independent of the active detector area, which is crucial for achieving superior noise performance, particularly during high count rates. To minimize stray capacitance and eliminate pick-up noise and microphony effects that would otherwise degrade spectroscopic performance, the first stage of electronics (internal n-JFET) is integrated directly onto the detector chip and connected to the anode through a small metal strip \cite{Niculae2006}.
SoLEXS uses PNDetector's SDD$^{\text{plus}}$ detector with a \SI{30}{\milli \metre \squared} active area, \SI{450}{\micro\metre} Silicon thickness, and \SI{8}{\micro\metre} of DuraBeryllium Plus window. The SDD$^{\text{plus}}$ series further reduces parasitic capacitance, enabling high-resolution and ultra-fast operations \cite{Niculae2013}, making it ideal for SoLEXS's requirement to process rapid solar X-ray flux variations during flare events.

Following charge collection, signal conditioning is performed through a Charge Sensitive Pre-Amplifier (CSPA), which converts the generated charge pulses into equivalent voltage signals. The CSPA output takes the form of a ramp signal with detector leakage current, where individual X-ray events appear as discrete steps superimposed on this ramp. To prevent saturation during periods of high solar activity, the ramp is periodically (\SI{2}{\milli\second}) discharged through reset pulses \cite{Niculae2006,Shanmugam2015,Bug2025}.

\begin{figure}
    \centering
    \includegraphics{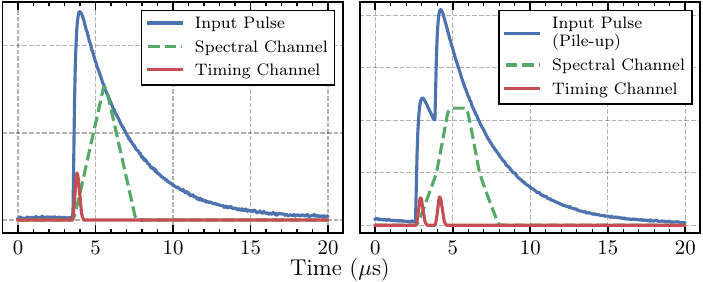}
    \caption{DPP demonstration for non-pileup (left panel) and pile-up (right panel) events. Each panel shows the input exponential pulse ($\tau$ = \SI{3.2}{\micro\second}) and the corresponding triangular-shaped outputs from the spectral channel (\SI{2}{\micro\second} peaking time) and the timing channel (\SI{.35}{\micro\second} peaking time, scaled $\times$\num{4} for visibility). In pile-up scenario, the faster timing channel resolves individual events that appear as single distorted pulses in the spectral channel.}
    \label{fig:dpp}
\end{figure}

The ramp output from the CSPA then undergoes amplification, filtering, and offset correction in an analog pre-filter card. This processing transforms the stepped ramp into discrete pulses corresponding to individual X-ray events, each characterized by a fast rise time and exponential decay with a time constant of \SI{3.2}{\micro\second} (Figure~\ref{fig:dpp}). This analog signals are then buffered and digitized through a 12-bit analog-to-digital converter (ADC) at \SI{20}{\mega\hertz} sampling rate. However, only the nine most significant bits (MSBs) are considered for further processing carried out in an FPGA.

The digitized signals are processed in FPGA using a Digital Pulse Processing (DPP) scheme to extract energy and timing information from incident X-ray photons. The fast-rising, exponentially-decaying digital pulse undergoes mathematical transformation to produce a triangular-shaped pulse (detailed processing blocks are described in Ref.~\citenum{Bug2025}). For each detector, two DPP chains run in parallel: a spectral (slow) chain and a timing (fast) chain. In these two chains, all the processing blocks are identical except for the width of the triangular-shaped pulse. The half of the triangular pulse width, defined as peaking time, is \SI{2}{\micro\second} for the spectral chain and \SI{.35}{\micro\second} for timing chain (Figure~\ref{fig:dpp}). 

An event is considered as valid X-ray photon event if the triangular pulse peaks in both spectral and timing chains exceed their respective threshold parameters, which are commandable parameters. 
The peak amplitude of valid triangular pulse of the spectral chain, a \SI{9}{\bit} number, is identified as a Pulse Height Analysis (PHA) channel, which is proportional to the energy of the corresponding incident X-ray photon. Every second, the number of pulses corresponding to each possible \num{512} channels are stored in the FPGA's built-in memory, generating a histogram that represents the X-ray energy spectrum for that interval.
Simultaneously, the timing chain counts valid triangular pulses every \SI{.1}{\second}, providing high-temporal-resolution light curves. Based on pulse amplitude, events are sorted into three coarse energy bins with commendable threshold limits.

During solar flares, the incident photon count rates can be very high (about \SIrange{1e5}{1e6}{\counts\per\second}) (Figure~\ref{fig:xrsb_solexs_scatter}). At such a high rate, the probability of two X-ray photons arriving at the detector within the spectral chain peaking time is relatively high. When such overlap occurs, the spectral chain's processed triangular pulse from the two photons ``pile-up" to produce a distorted triangular pulse (see right panel of Figure~\ref{fig:dpp}). In such a case, the peak of the distorted triangular pulse is wrongly stored as a valid PHA channel. The resultant energy spectrum is thus highly distorted at high incident count rates.
To prevent the recording of such pile-up events and maintain the accuracy of the energy spectrum, pile-up rejection (PUR) logic is included in the pulse processing. The timing chain processes the X-ray event faster, and the two pile-up events that are not resolved by the spectral chain are resolved by the timing chain (see right panel of Figure~\ref{fig:dpp}). Two photons piling up in the timing chain is therefore a rarer occurrence compared to the spectral chain. If multiple events are detected by the timing channel during the processing of a single spectroscopic pulse, the event is considered a pile-up, and the corresponding pile-up PHA channel is not be stored as a valid event. However, the spectral pile-up events are individually recorded by the faster timing chain and are stored as valid counts. Hence, the number of recorded events in the timing chain is much more representative to the actual count rate compared to the spectral count rate. 

\section{Ground Calibration} \label{sect:ground_cal}
The raw spectral and temporal data from SoLEXS do not directly represent the incident solar X-ray spectra. Instrumental effects, such as the energy-channel relationship, energy-resolution relationship, instrumental response, deadtime and pile-up, systematically affect these measurements. Calibration of the instrument is essential for the characterisation of these instrumental effects to accurately and reliably convert the raw signals into science data.
	
Ground calibration involves systematic characterisation of the instrument's fundamental performance parameters. The ground calibration establishes the energy-channel relationship and energy-resolution relationship across the operational energy range using well-characterised radioactive sources, primarily $^{55}$Fe and $^{241}$Am, along with X-ray fluorescence targets that provide characteristic line emissions. These measurements determine the gain and offset parameters essential for accurate energy determination of detected photons. 

Instrumental response characterisation forms a critical component of the calibration process, involving detailed modelling of the instrument's response to incident X-rays across different energies. This includes mapping the detector's effective area, quantum efficiency variations, and the spectral redistribution function. Collimator response measurements further characterise the instrument's field of view and angular response to ensure accurate flux determinations, while deadtime characterisation involves systematic assessment of the detector's count rate capability.

The derived calibration parameters are validated through thermo-vacuum performance testing conducted under simulated space conditions. These tests verify the stability and accuracy of calibration parameters across the operational temperature range. The following subsections detail each calibration procedure and its implementation.

\subsection{Energy-Channel Calibration}

\begin{figure}
    \centering
    \includegraphics{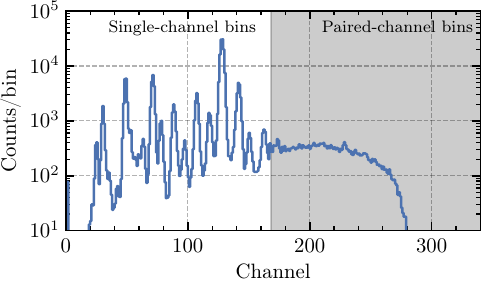}
    \caption{Sample XRF spectrum of JSC-1A lunar simulant with salt obtained using SDD1, demonstrating the instrument's \num{340}-channel binning scheme. Channels up to \num{168} (left region) represent single-channel binning, while channels \numrange{169}{340} (right, shaded region) show paired-channel binning. The absence of signal above channel \num{280} 
    results from the X-ray gun operating voltage of \SI{20}{\kilo\volt}, which limits the maximum photon energy available.}
    \label{fig:jsc-1a_channel_bins}
\end{figure}

SoLEXS's DPP system converts the digitized signals from the detector into shaped triangular pulses (see Section~\ref{sect:dpp}). The peak of each pulse is recorded as a PHA channel, representing the photon energy. The PHA channels are initially recorded using \SI{9}{\bit}, creating \num{512} possible channels. 
An adaptive energy binning is implemented by maintaining \SI{\approx 47.75}{\electronvolt} channel width up to channel \num{168} (approximately \SI{8}{\keV}), which contains the majority of solar coronal emission lines \cite{Phillips2004}, then grouping subsequent channels into pairs, doubling the channel width to \SI{\approx 94.5}{\electronvolt} for higher energies (Figure~\ref{fig:jsc-1a_channel_bins}). 
This approach preserves finer binning for diagnostics of soft X-ray lines while accommodating broader continuum measurements up to \SI{22}{keV}. 
The resulting \num{340}-channel configuration balances scientific requirements while minimizing data volume, providing sufficient spectral resolution for temperature and abundance diagnostics.

The relationship between incident X-ray photon energy and the PHA channel is governed by two parameters: gain, which defines the proportionality factor between photon energy and the PHA channel and determines the energy bin width, and offset, which determines the baseline channel value. 
	
\begin{figure}[h!]
    \centering
    \includegraphics{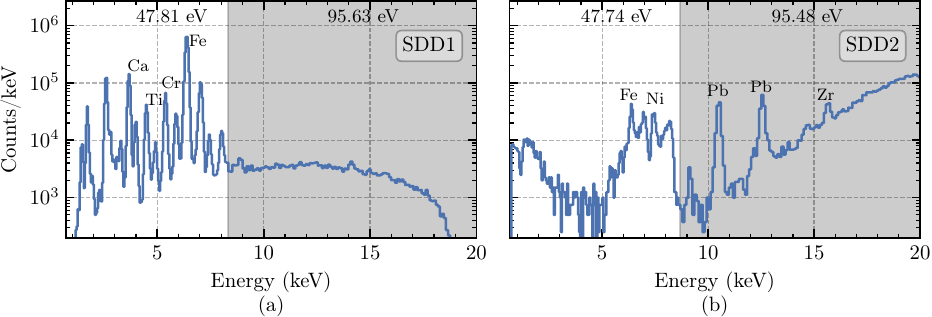}
    \caption{XRF spectrum of JSC-1A lunar simulant with salt obtained using SDD1 (a) and Mu metal obtained using SDD2 (b), demonstrating the energy-channel calibration methodology. The spectrum exhibits characteristic fluorescence lines which serve as reference energies for establishing energy-channel relationship. 
    The absence of signal above $\approx$\SI{19}{\keV} in SDD1 spectrum (a) results from the X-ray gun operating voltage of \SI{20}{\kilo\volt}, which limits the maximum photon energy available.
    The SDD2 spectrum (b) below 5 keV is significantly attenuated due to absorption by the Mu metal target.        
    }
    \label{fig:sdd12_spec}
\end{figure}

For energy-channel calibration, characteristic X-ray fluorescence (XRF) lines are generated using silver-anode X-ray gun to irradiate XRF targets. The primary X-rays from the gun eject inner shell electrons from target atoms, causing outer shell electrons to cascade down and emit characteristic fluorescence lines specific to each element present in the target material. Additionally, the X-ray gun produces a continuum spectrum through bremsstrahlung processes that underlies the discrete fluorescence lines (Figure~\ref{fig:sdd12_spec}).

SDD1 calibration utilises the JSC-1A lunar regolith simulant \cite{jsc1a_cite} mixed with salt to generate XRF lines of Calcium, Titanium, Chromium, and Iron as illustrated in Figure~\ref{fig:sdd12_spec}a. SDD2 calibration utilizes Mu metal \cite{Smith1927} targets to generate Iron and Nickel XRF lines, supplemented by Lead lines originating from PCB solder and Zirconium lines from the detector's internal collimator, as illustrated in Figure~\ref{fig:sdd12_spec}b. The experimental setup is described in Section~\ref{section:thermovac}. 

\begin{figure}[h!]
    \centering
    \includegraphics{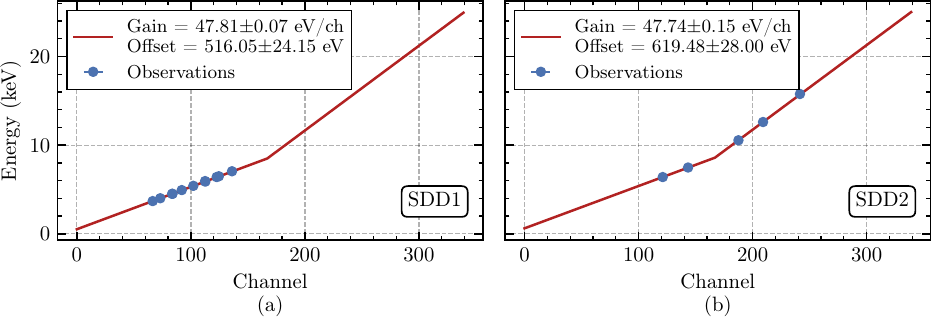}
    \caption{Energy-channel calibration plots for SDD1 (a) and SDD2 (b) displaying the linear relationship between incident photon energy and recorded PHA channel. The plots demonstrate the instrument's energy binning scheme, with single-channel binning up to \num{168} channel, followed by paired-channel binning that doubles the energy width for channels \numrange{169}{340}, as indicated by the change in slope after the 168th channel. }
    \label{fig:sdd12_e_ch}
\end{figure}

The observed centroid channels of these XRF lines are used to establish the linear relationship between energy and channel number for each detector. The resulting energy-channel calibration and derived gain and offset parameters are presented in Figure~\ref{fig:sdd12_e_ch} for SDD1 and SDD2. The temperature dependence of these calibration parameters is systematically characterized during comprehensive thermo-vacuum testing.

\subsection{Energy-Spectral Resolution Calibration}\label{section:energy_resolution}
The energy resolution of a spectroscopic instrument characterizes its ability to discriminate between adjacent photon energies.
In SDDs, this resolution is governed by fluctuations and uncertainties in charge collection and pulse processing chain that converts incident photon energy into discrete PHA channels. 
	
Total spectral broadening arises from two primary sources. Fano noise represents the fundamental limit due to statistical variations in electron-hole pair creation, where the number of charge carriers scales linearly with photon energy  ($N=E/w$, where $w$=\SI{3.62}{\electronvolt} for Silicon \cite{Klein1968}). Electronic noise originates from the readout electronics and contributes a fixed energy-independent broadening component. The quadrature combination of these noise sources determines the overall energy resolution according to the resolution model given in Equation~\ref{eq:fwhm}\cite{knoll2010radiation}.

\begin{equation}\label{eq:fwhm}
\text{FWHM}(E) = \sqrt{\text{FWHM}_{\text{EN}}^2 + 8\cdot \ln{2}\cdot F \cdot w \cdot E},
\end{equation}

where FWHM$_{\text{EN}}$ represents the electronic noise contribution, $F$ is the Fano factor for Silicon (\num{0.129}) \cite{Kotov2018}, and $E$ is the incident photon energy.

\begin{figure}
    \centering
    \includegraphics{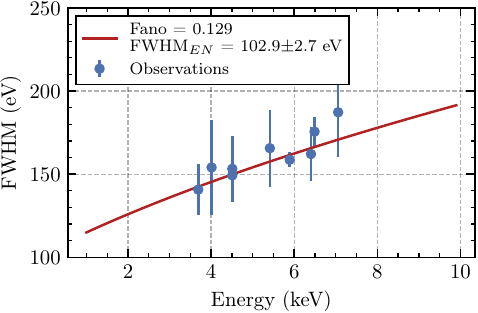}
    \caption{Observed spectral resolution at different energies (blue data points) over-plotted with resolution model from Equation~\ref{eq:fwhm} (red solid curve).}
    \label{fig:e_fwhm}
\end{figure}

The energy resolution calibration is accomplished using X-ray fluorescence (XRF) and radioactive source lines spanning the instrument's operational energy range. The energies employed include Fe K$\alpha$ and K$\beta$, Ti K$\alpha$, Cr K$\alpha$, Ca K$\alpha$ and K$\beta$ lines generated from XRF of JSC-1A regolith simulant mixed with salt, as well as Mn K$\alpha$ and K$\beta$ lines from the on board $^{55}$Fe radioactive calibration source. 
The main photopeak is fitted with a Gaussian profile for each calibration line to extract the full width at half maximum (FWHM), which are subsequently fitted to the resolution model with the Fano factor fixed at the theoretical Silicon value. The best-fit electronic noise contribution is determined to be \SI{102.9 \pm 2.7}{\electronvolt} (FWHM). Figure~\ref{fig:e_fwhm} demonstrates the energy resolution performance across the energy range, showing FWHM as a function of incident photon energy fitted with the resolution model (Equation~\ref{eq:fwhm}).

\subsection{Thermo-vacuum Test}\label{section:thermovac}
\begin{figure}
    \centering
    \includegraphics[width=0.6\linewidth]{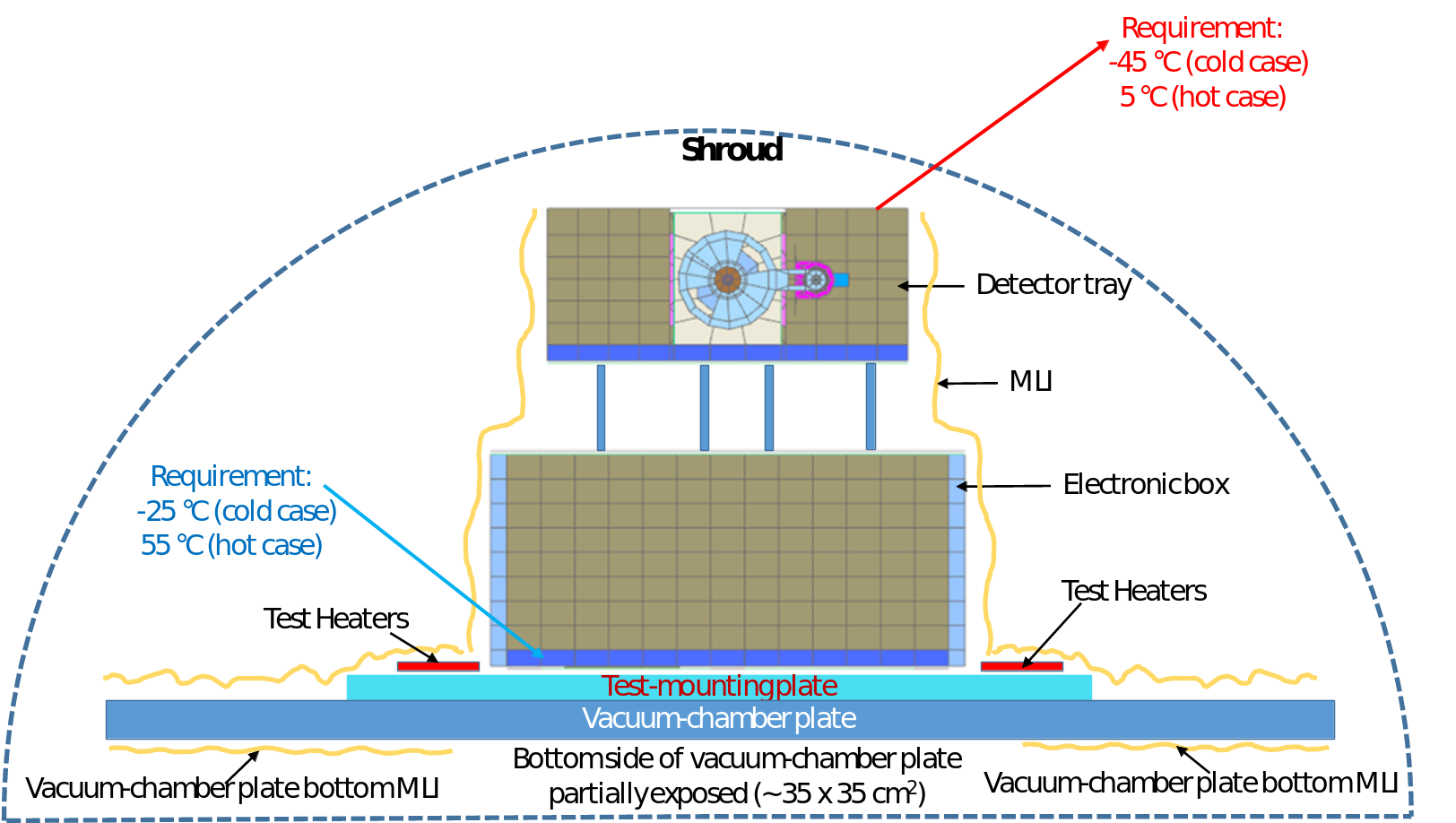}
    \caption{TVP test setup. The flight configured payload including MLI was bolted on to the test-mounting plate of \SI{6}{\milli\metre} thickness (\SI{0.5}{\metre}$\times$\SI{0.5}{\metre}). The test-mounting plate was bolted to the vacuum-chamber plate (Aluminium \SI{10}{\milli\metre} thick) and was equipped with test heaters and thermocouples. The vacuum chamber shroud was maintained between \SI{-35}{\celsius} and \SI{-100}{\celsius} to simulate the deep space radiative environment. }
    \label{fig:tvac_setup}
\end{figure}

The thermo-vacuum performance (TVP) test is conducted to characterize the temperature response of gain, offset, and resolution parameters under simulated operational environment. 
The TVP test is carried out at Environment Test Facility at URSC using a \SI{1.6}{\metre} diameter vacuum chamber. The flight-configured SoLEXS payload is mounted onto a temperature-controlled Aluminium plate using flight-equivalent thermal interface materials, with the assembly subsequently bolted to the vacuum chamber's Aluminium base plate (Figure~\ref{fig:tvac_setup}). Strategic placement of test heaters and thermocouples throughout the mounting configuration enabled precise thermal control and comprehensive temperature monitoring during all test phases. The instrument was in flight configuration including multi-layer insulation (MLI), with the exception that the aperture cover mechanism was not integrated to enable detector calibration using external X-ray sources during thermal cycling.

\begin{figure}
    \centering
    \includegraphics[width=0.9\linewidth]{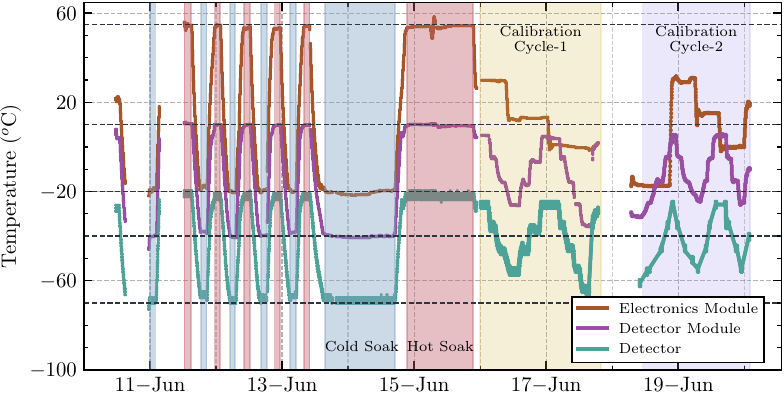}
    \caption{Temperature cycling during TVP test conducted from 10 June to 20 June 2022, showing the thermal qualification and calibration phases. 
    The temperature ranges corresponded to the environmental test level specification, with detector module temperatures cycling between \SI{-40}{\celsius} and \SI{0}{\celsius}, and electronics module temperatures between \SI{-15}{\celsius} and \SI{50}{\celsius}.
    The detector temperatures are approximately \SI{30}{\celsius} below the detector module temperature through active Peltier cooling. Two calibration cycles were performed with systematic variation of detector and electronics module temperatures to characterize the thermal dependencies of gain and offset parameters.
    }
    \label{fig:tvac_cycle}
\end{figure}

\begin{figure}
    \centering
    \includegraphics{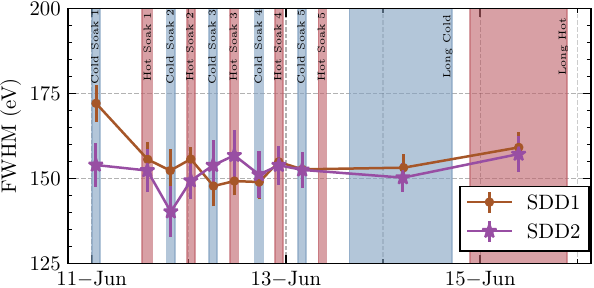}
    \caption{Energy resolution (FWHM) at Mn K$\alpha$ (\SI{5.9}{\keV}) measured during thermal qualification cycling of the TVP test for both SDD1 (circles) and SDD2 (asterisks) detectors using the internal $^{55}$Fe calibration source. The energy resolution remained stable within \SI{\pm 15}{\electronvolt} across the qualification temperature range.}
    \label{fig:tvac_fwhm}
\end{figure}

The thermal cycle consisted of five short hot and cold soaks, each lasting two hours, followed by extended hot and cold soaks of 24 hours each (Figure~\ref{fig:tvac_cycle}). 
Throughout these thermal cycles, SDDs' performance were monitored in terms of energy resolution at Mn K$\alpha$, using the internal calibration source. The results show that the energy resolution remained constant (within \SI{\pm 15}{\electronvolt}) across the qualification temperature range (Figure~\ref{fig:tvac_fwhm}). 
	
Following the thermal cycling qualification, two calibration cycles were performed with systematic variation of detector and electronics module temperatures to characterize the thermal dependencies of gain and offset parameters. In the first calibration cycle, SDD1 is calibrated using X-ray fluorescence from JSC-1A regolith simulant mixed with salt pellet, with the X-ray gun operated in reflective geometry.
Due to its smaller aperture area and higher flux requirements, the second calibration cycle for SDD2 employed X-ray fluorescence from Mu metal, with the X-ray gun in transmission geometry. 
	
\begin{figure}
    \centering

    \begin{subfigure}[b]{0.48\textwidth}
        \centering
        \includegraphics[width=\textwidth]{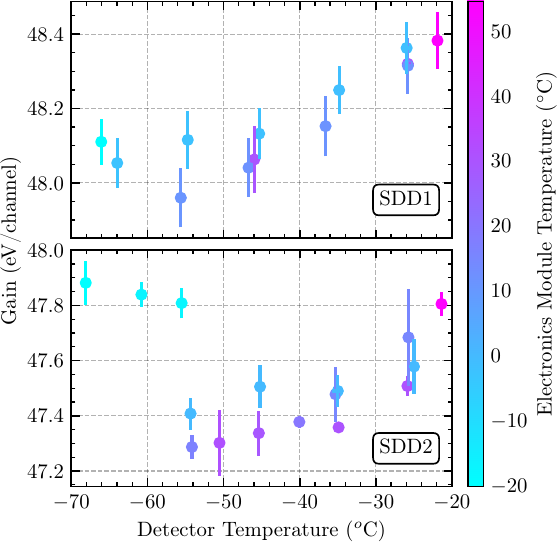}
        \caption{Gain dependence on detector temperature.}
        \label{fig:tvac_sdd12_gain}
    \end{subfigure}
    \hfill 
    \begin{subfigure}[b]{0.48\textwidth}
        \centering
        \includegraphics[width=\textwidth]{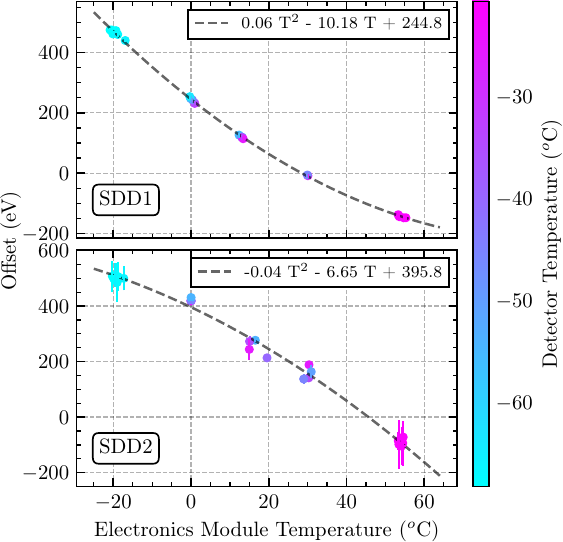}
        \caption{Offset dependence on electronics temperature.}
        \label{fig:tvac_sdd12_offset}
    \end{subfigure}

    \caption{Thermal dependence of SoLEXS calibration parameters during thermo-vacuum cycles for SDD1 (top panel in each plot) and SDD2 (bottom panel in each plot). 
    (a) Gain exhibits a weak dependence on detector temperature, with points color-coded by electronics module temperature. 
    (b) Offset shows a clear quadratic dependence on electronics module temperature, with points color-coded by detector module temperature.}
    \label{fig:tvac_gain_offset_combined}
\end{figure}

The analysis reveals that the gain parameter weakly depends only on the detector temperature (Figure~\ref{fig:tvac_sdd12_gain}) while the offset parameter depends only on the electronics module temperature (Figures~\ref{fig:tvac_sdd12_offset}). These findings are used for temperature dependent correction during orbital operations.

\subsection{Instrument Response}\label{section:instrument_response}

The temperature-dependent gain and offset parameters, established during TVP testing, enable conversion of observed spectrum in PHA channels into energy space. However, the resulting energy space spectrum does not directly represent the incident X-ray spectrum due to instrumental response that modify the detected photon distribution.
	
The transformation from incident to observed spectra is governed by two components that collectively define the instrument's spectral response: the Spectral Redistribution Function (SRF) and the Ancillary Response Function (ARF). 
The SRF characterizes the detector's energy redistribution, while the ARF accounts for the instrument's geometric area and energy-dependent efficiency. Mathematically, the observed spectrum $P(c)$ is convolution of the incident photon spectrum $F(E)$, modified by the ARF $A(E)$, with SRF $R(c,E)$ \cite{xspec_guide},

\begin{equation} \label{eq:response}
P(c) = \int F(E) \cdot A(E) \cdot R(c, E) \, dE ,
\end{equation}

where c is instrument PHA channels, and E is incident photon energy. The spectral response $R(c,E_0)$ at a specific energy $E_0$ represents how the mono-energetic input spectrum of energy $E_0$ is redistributed by the instrument in PHA channels. The ancillary response $A(E_0)$ quantifies effective area at some energy $E_0$. The knowledge of both the SRF and ARF enables the transformation required to recover incident photon spectra $F(E)$ from observed detector measurements $P(c)$.

\subsubsection{Spectral Redistribution Function}

\begin{figure}
    \centering
    \includegraphics{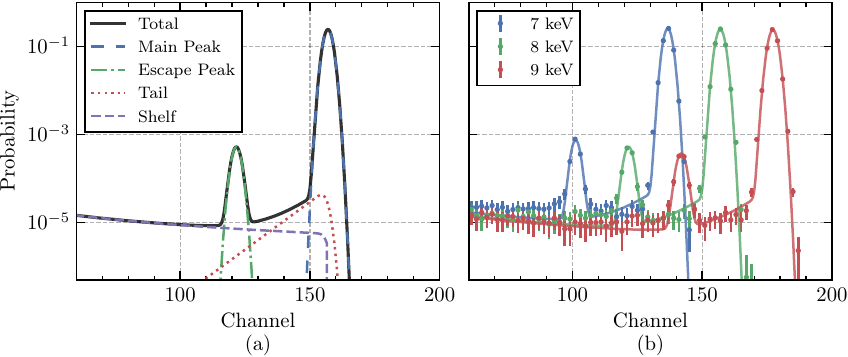}
    \caption{SoLEXS spectral response characterization using HYPERMET modeling. (a) Four characteristic components of the HYPERMET SRF: main peak, Silicon escape peak (\SI{1.74}{\keV} below main peak), exponential tail, and low-energy shelf. (b) SoLEXS spectra (data points) obtained with monoenergetic synchrotron radiation at \num{7}, \num{8}, and \SI{9}{\keV} from RRCAT BL-16 beamline, with fitted HYPERMET models (solid lines) overlaid.}
    \label{fig:rrcat}
\end{figure}

The SRF quantifies how mono-energetic photons are redistributed across PHA channels due to detector physics and processing electronics. The resulting response profile is not a simple Dirac-delta response but a sum of four distinct components:
a main Gaussian peak from events with complete charge collection,
a Silicon escape peak located (\SI{1.74}{\keV}) below the main peak, caused by escaping Si-K$\alpha$ fluorescence photons,
an exponential low-energy tail resulting from incomplete charge collection.
The complete SRF is expressed as (Figure~\ref{fig:rrcat}a):
\begin{equation}
R(c, E) = \text{Main}(c) + \text{Escape}(c) + \text{Tail}(c) + \text{Shelf}(c),
\end{equation}
where c is instrument PHA channels, and E is incident photon energy. 
Each of these four components is modeled mathematically using the HYPERMET function framework \cite{Phillips1976} as follows:
\newline
1. Main Gaussian peak
\begin{equation}
\frac{I_{\text{main}}}{\sqrt{2\pi\sigma_m^2}} \exp\left( -\frac{(c - c_m)^2}{2\sigma_m^2} \right),
\end{equation}
2. Silicon escape peak 
\begin{equation}
\frac{I_{\text{esc}}}{\sqrt{2\pi\sigma_e^2}} \exp\left( -\frac{(c - c_e)^2}{2\sigma_e^2} \right),
\end{equation}
3. Tail component (empirical, exponential + erfc) 
\begin{equation}
I_{\text{tail}} \exp\left( \frac{c - c_m}{\beta} \right) \, \text{erfc}\left( \frac{c - c_m}{\sqrt{2}\sigma} + \frac{\sqrt{2}\sigma}{2\beta} \right),
\end{equation}
4. Shelf component (empirical, for $c < c_m$) 
\begin{equation}
\frac{I_{\text{shelf}}}{2} \left( \frac{c}{10} \right)^{-\alpha} \text{erfc}\left( \frac{c - c_m}{\sqrt{2}\sigma} \right), \quad \text{for } c < c_m.
\end{equation}

Here, $c_m$ and $c_e$ are the centroid channels of the main and escape peaks, respectively.
The term $\sigma$ represents the standard deviation (width) of the Gaussian components ($\sigma_m$ for the main peak, $\sigma_e$ for the escape peak), which is determined by the combined Fano and electronic noise as described in Section~\ref{section:energy_resolution}. 
$I_{\text{main}}$, $I_{\text{esc}}$, and $I_{\text{tail}}$ are the relative intensities of the corresponding spectral components.
Finally, $\alpha$ and $\beta$ are empirical parameters with $\alpha$ controlling the power-law slope of the shelf component arising from electron escape events and $\beta$ governing the exponential decay of the tail caused by incomplete charge collection.

Experiments to characterize the SRF were conducted at the BL-16 beamline of Raja Ramanna Centre for Advanced Technology (RRCAT), Indore (\url{https://www.rrcat.gov.in}), using mono-energetic X-rays (\SIrange{6.5}{16}{\keV}) from a synchrotron source. Beam energy is tuned using a dual-crystal monochromator, and flux is regulated to \SIrange{1000}{2000}{\counts\per\second} via adjustable slits. 
	
The acquired monochromatic spectra are fitted using HYPERMET model to derive the parameters
and validated through representative spectra acquired at three characteristic energies as shown in Figure~\ref{fig:rrcat}b.
The energy dependence of each parameter is derived empirically.
While the energy-dependent behaviour of $\sigma$ follows the relationship described in Section~\ref{section:energy_resolution}, the relative component intensities exhibit exponential energy dependencies, whereas $\alpha$ and $\beta$ remains constant across the measurement range. 

The empirical relationships governing the intensity ratios of escape-to-main-peak and tail-to-main-peak are: $\frac{I_{\text{esc}}}{I_{\text{main}}} = \exp(-4.54 - 0.20 E)$ and $\frac{I_{\text{tail}}}{I_{\text{main}}} = \exp(-13.18 - 0.34 E)$.
These empirical relationships are used to generate the complete HYPERMET SRF across the entire energy range through extrapolation beyond the experimental calibration energies.
Experimentally, the main peak's response probability exceeds \SI{99}{\percent} at \SI{6}{\keV}, with all other spectral components contributing less than \SI{1}{\percent}.
Due to dominant main peak contribution, simplified Gaussian SRF model containing only the main peak and escape peak components are used for spectral fitting during high solar activity.
However, during solar minimum conditions with lower count rates and extended integration times, the complete non-Gaussian HYPERMET SRF model will be used. 

\subsubsection{Ancillary Response Function}
The theoretical effective area incorporates the detector's geometric and material properties combined with the geometric aperture area ($A_{\mathrm{geo}}$). 
The calculation assumes SDD with \SI{8}{\micro\metre} Moxtek DuraBeryllium Plus entrance window, \SI{10}{\nano\metre} dead Silicon layer, and \SI{450}{\micro\metre} active detector thickness. 
The energy-dependent transmission efficiency is calculated using Beer-Lambert attenuation through successive material layers:
\begin{equation}
A(E) = A_{\mathrm{geo}} \times T_{\mathrm{win}}(E) \times
\exp\left(-\sigma_{\mathrm{Si}}(E), \rho_{\mathrm{Si}}, t_{\mathrm{Si,dead}}\right) \times
\left[1 - \exp\left(-\sigma_{\mathrm{Si}}(E), \rho_{\mathrm{Si}}, t_{\mathrm{Si,active}}\right)\right],
\label{eq:arf_updated}
\end{equation}
where $T_{\mathrm{win}}(E)$ is the energy-dependent transmission of the DuraBeryllium Plus window, with values taken directly from the manufacturer's datasheet\footnotemark[2], $\sigma(E)$ represents energy-dependent absorption cross-sections of the corresponding materials taken from Ref.~\citenum{Elam2002}, $\rho$ denotes the corresponding material density, and $t$ is the material thickness. Figure~\ref{fig:theo_arf} presents the calculated theoretical ancillary response functions for both SoLEXS detectors.

\subsection{Deadtime Correction}

Accurate determination of incident photon fluxes, especially during high-rate solar flares, requires correction for instrument deadtime. This phenomenon, where the system's finite processing time ($\tau$) causes subsequent events to be missed, must be characterized to convert observed count rates to true photon fluxes.

In SoLEXS, deadtime primarily arises due to the DPP system, with additional contribution from PUR logic, which extends $\tau$ to avoid pile-up events. 
A detector emulator was used to inject simulated pulses to characterise the pulse processing deadtime, mimicking the fast rise and \SI{3.2}{\micro\second} exponential decay of the SDD output into the DPP system. Input rates (Poisson distributed to represent source distribution) were varied from \SIrange{1e4}{1e5}{\counts\per\second}. The recorded output rates from the spectral (slow) and temporal (fast) chains are compared to the known inputs to quantify the deadtime.

\begin{figure}[h!]
    \centering
    \begin{minipage}[t]{0.48\textwidth}
        \centering
        \includegraphics[width=\textwidth]{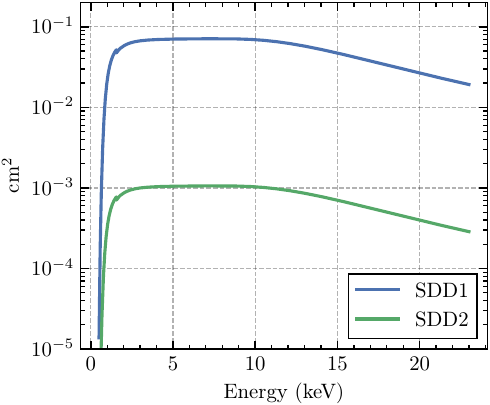}
        \caption{Theoretical ARFs for both SoLEXS detectors, calculated using Beer-Lambert attenuation. The measured geometric aperture areas are \SI{7.1063 \pm 0.0095}{\milli\metre\squared} and \SI{0.1065 \pm 0.0006}{\milli\metre\squared} for the large (SDD1) and small (SDD2) apertures respectively.}
        \label{fig:theo_arf}
    \end{minipage}
    \hfill 
    \begin{minipage}[t]{0.48\textwidth}
        \centering
        \includegraphics[width=\textwidth]{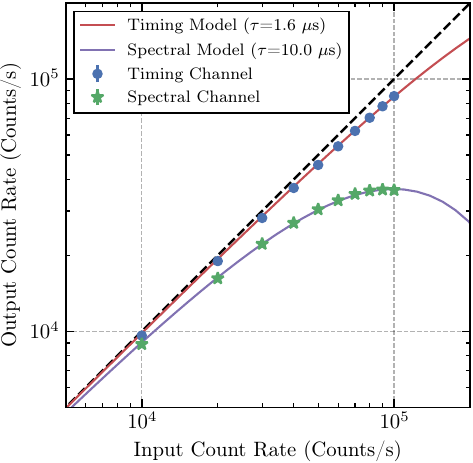}
        \caption{Recorded output count rate plotted against input count rate for timing chain and spectral chain. The observations were fitted with paralyzable deadtime model.}
        \label{fig:pile_up_model}
    \end{minipage}
    
\end{figure}

The results show that the SoLEXS's DPP system follows an idealised paralyzable deadtime model. Hence, the observed (measured) count rate, assuming Poisson distributed incident events, $m$ is given by \cite{Usman2018,knoll2010radiation} $m = n \cdot e^{-n\cdot\tau}$ (where $\tau$ is deadtime and n is the true count rate).
From these calibration experiments, the deadtime for the spectral chain is determined to be \SI{10}{\micro\second}, while the temporal chain exhibits a much shorter deadtime of \SI{1.6}{\micro\second} (Figure ~\ref{fig:pile_up_model}). The spectral pile-up is only \SI{\sim 1}{\percent} even at the high count rate of $10^5$ counts/second, owing to the PUR logic.

At high incident rates (\SI{>5e4}{\counts\per\second}), the spectral chain's measured count rate saturates and starts declining due to the paralyzable nature of its deadtime. This degeneracy creates ambiguity in determining the actual incident rate. In contrast, the temporal chain's shorter deadtime preserves a one-to-one relationship between measured and true rates even at \SI{\sim 1e5}{\counts\per\second}. Thus, the measured temporal chain's rate can be inverted using a look-up table to obtain the actual rate.

\subsection{Collimator Response}

The measured X-ray flux undergoes systematic modifications due to ARF and deadtime effects as discussed in previous sections, with additional modification arising from the position of source within the FOVs, hence, the angular response of two cylindrical apertures. 
In order to estimate the same, experiments were conducted using an X-ray gun with a copper anode mounted on a computer-controlled two-axis translation stage positioned \SI{2.5}{\metre} from the aperture plane.

The detector module assembly, containing a \textit{Ketek} SDD 
identical to the flight units and positioned at the precise location of the flight detectors, was mounted on a flat, height-adjustable platform with its mechanical  axis aligned with the X-ray gun's axis. A raster scan of the X-ray gun was performed in increments of \SI{0.1}{\degree} across a grid spanning \SI{\pm 2}{\degree} relative to the both aperture axes, covering the full angular acceptance range. At each grid position, spectral data was acquired, and the \SI{8.05}{\keV} Cu K$\alpha$ line was used to generate the experimental two-dimensional collimator response profile across the FOVs.
	
\begin{figure}[h!]
    \centering
    
    \begin{minipage}[t]{0.48\textwidth}
        \centering
        \includegraphics[width=\textwidth]{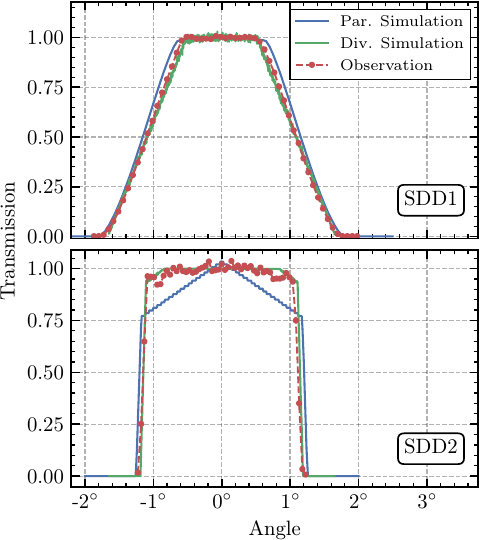}
    \caption{Collimator response measurements for SDD1 (top panel) and SDD2 (bottom panel), with experimental data points (red) obtained using an X-ray gun performing raster scans across a \SI{\pm 2}{\degree} grid. The agreement between divergent ray simulations (green) and experimental measurements validates the geometric parameters of the cylindrical apertures, while the parallel ray simulation (blue) provides the angular correction function for on-orbit solar observations.}
        \label{fig:sdd12_collimator}
    \end{minipage}
    \hfill 
    \begin{minipage}[t]{0.48\textwidth}
        \centering
        \includegraphics[width=\textwidth]{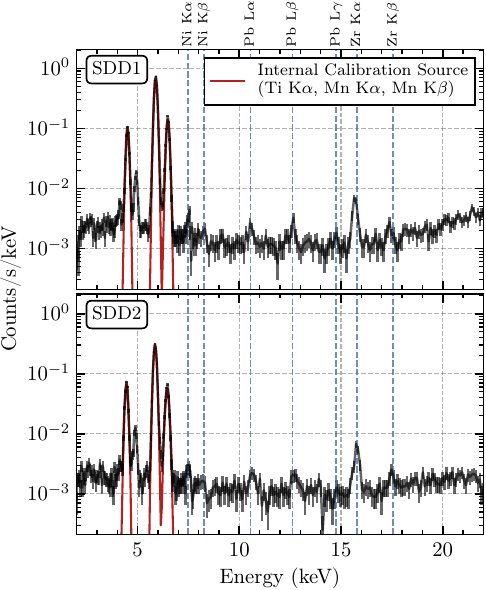}
    \caption{On board calibration spectra from SDD1 (top panel) and SDD2 (bottom panel) acquired during the pre-commissioning phase over \num{48} hours using the internal calibration source. Additional instrumental fluorescence lines from Nickel, Lead, and Zirconium are evident in the spectra. 
    Gaussian model (solid curves) is fitted to the calibration source spectra to determine spectral resolution at Mn K$\alpha$.
    The slight differences in relative intensities between the two detectors arise from slight asymmetry in calibration source orientation, consistent with the ground calibration observations.
    }
        \label{fig:pre_commi_sdd12}
    \end{minipage}
    
\end{figure}

In order to validate the experimental results, simulations are performed under two conditions: one assuming perfectly parallel rays (representative of solar illumination) and another with divergent rays (representative of practical experimental conditions accounting for the finite source distance) shown in Figure~\ref{fig:sdd12_collimator}. By comparing the divergent-ray simulation and the experimental collimator response, the inner and outer radii of the cylindrical apertures are verified against independent profile projector measurements. The final parallel-ray collimator responses, derived from these geometric parameters and representative of actual collimated solar illuminations, are used as the angular response corrections.

\section{On board Calibration: Post Launch } \label{sect:onboard_cal}
	
\subsection{Pre-Commissioning Phase}
Following the launch on 2 September 2023 and subsequent cruise phase to L1, once the Aditya-L1 spacecraft entered a stable thermal and radiation environment, SoLEXS was powered on for the first time in space on 16 October 2023. The initial power on occurred with the SMA mechanism in the closed condition, and the detectors were not exposed to direct solar X-rays. In this configuration, the internal calibration source 
was used to assess the instrument's baseline performance under flight conditions, using characteristic Mn and Ti K$\alpha$, K$\beta$. 
During this phase, the detector module was stabilised at a temperature of \SI{-40 \pm 3}{\celsius}, with the temperature difference between the detector module and the detector maintained at approximately \SI{30}{\celsius}. The electronics module was operated at a temperature of \SI{40 \pm 2}{\celsius}. 

Figure~\ref{fig:pre_commi_sdd12} presents the spectra acquired from the internal calibration source over an integration period of \num{48} hours for SDD1 and SDD2. The full width at half maximum (FWHM) at the Mn K$\alpha$ line was measured to be \SI{164.9 \pm 4.7}{\electronvolt} for SDD1 and \SI{171.2 \pm 5.9}{\electronvolt} for SDD2, with corresponding gains of \SI{48.24 \pm 0.28}{\electronvolt\per\channel} (SDD1) and \SI{47.52 \pm 0.33}{\electronvolt\per\channel} (SDD2), and offsets of \SI{-33.88 \pm 33.44}{\electronvolt} (SDD1) and \SI{86.66 \pm 39.21}{\electronvolt} (SDD2). These values closely match those obtained during ground calibration at corresponding operating temperatures.
Upon switch-on in closed aperture condition, a low-energy noise peak was observed in the spectra, requiring an increase in the spectral and temporal thresholds to approximately \SI{2}{\keV}.

\subsection{Performance Verification Phase}
\begin{figure}
    \begin{minipage}[t]{0.48\textwidth}
        \centering
        \includegraphics[width=\textwidth]{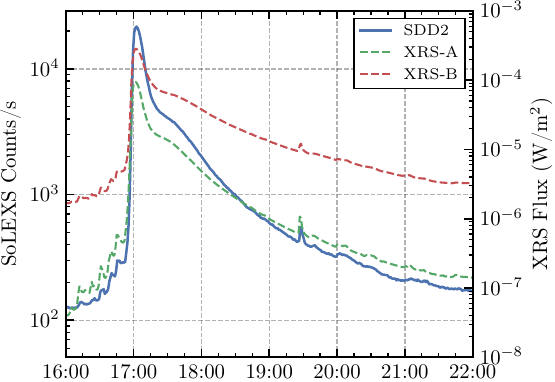}
        \caption{Light curve of the first X-class solar flare (X2.9, SOL2023-12-14T17:02) observed by SoLEXS, detected within 33 hours of aperture cover deployment. The solid blue curve shows SDD2 count rates (left y-axis) overlaid with GOES XRS-A (green dashed) and XRS-B (red dashed) flux measurements (right y-axis).}
        \label{fig:first_lc}
    \end{minipage}
    \hfill 
    \begin{minipage}[t]{0.48\textwidth}
        \centering
        \includegraphics[width=\textwidth]{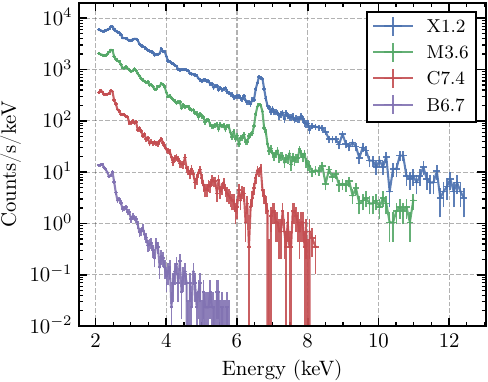}
        \caption{Representative solar flare spectra observed during flare peak by SDD2 during the PV phase. The spectra demonstrate the characteristic thermal continuum and emission line features across different flare intensities: X1.2-class (SOL2024-05-14T12:55, blue), M3.6-class (SOL2024-04-23T03:19, green), C7.4-class (SOL2024-04-23T21:11, red), and B6.7-class (SOL2024-04-06T21:26, purple).}
        \label{fig:flare_spectra}
    \end{minipage}
\end{figure}

The SMA mechanism was successfully deployed on 13 December 2023 (08:00 UTC), enabling direct solar X-ray observations and marking the commencement of the performance verification (PV) phase, which continued through June 2024. Within 33 hours of the aperture opening, SoLEXS detected an X2.9-class flare (SOL2023-12-14T17:02), the most intense solar event since September 2017 up to that date (Figure~\ref{fig:first_lc}). SDD2 (\SI{0.1065}{\milli\metre\squared} aperture) became the primary detector due to solar activity being at maximum levels. SDD1 (\SI{7.1063}{\milli\metre\squared}), optimised for solar minima conditions, saturated at flare-associated count rates exceeding \SI{1e5}{\counts\per\second}
and thus remained in a minimal operation mode during this phase.

As a result, a full on board characterization of SDD1 using direct solar flux is still pending.
While SDD1's ground and pre-commissioning phase calibrations are complete, and its key performance parameters, such as its gain, offset, and spectral resolution (FWHM), continue to be successfully monitored using the internal calibration source during off-pointing maneuvers. The major remaining task is the on board characterization of its deadtime corrections, which requires observing a range of solar fluxes without saturation. This final calibration step will be undertaken once the solar background activity subsides to A/B-class levels.

While full on board calibration of SDD1 was deferred, the PV phase was used to complete the on board characterisation of SDD2.
The PV phase coincided with a period of rapidly evolving solar activity: while April 2024 was characterised by relatively low X-ray background levels reaching $\sim$ B5 levels, May 2024 became the most active month of Solar Cycle 25 with the cycle's second strongest flare measuring X8.7 (SOL2024-05-14T16:51 UTC). 
Representative flare spectra captured during this period demonstrate the capability to observe the complete range of solar activity from B-class background events to X-class major flares (Figure~\ref{fig:flare_spectra}).
This pronounced variability in solar X-ray flux provided a dataset for evaluating the instrument's response and stability across its wide dynamic range.
The PV phase is utilised for refining calibration parameters, analysing deadtime effects, validating cross-calibration with external instruments, and assessing spectral response for SDD2.

\subsection{On Board Deadtime Correction}
During on board operations, it is observed that the reset pulse generated in the CSPA card, after passing through the analog pre-filter, produces spurious ringing-like signals. These signals are picked up by the DPP's timing chain and erroneously processed as valid events (similar observations by SphinX \cite[Section 5]{Gburek2012}). Since reset pulses are applied periodically every \SI{2}{\milli\second}, this results in a background of approximately \num{500} spurious counts per second in the timing chain. However, these spurious signals fall outside the bandwidth of the spectral chain and are not detected there.

When a genuine X-ray event coincides with a spurious reset pulse within the spectral deadtime, the timing chain registers both the genuine and spurious events, while the spectral chain records only the genuine event. However, due to the PUR logic, the genuine X-ray event is rejected if it coincides with the reset pulse, reducing the detection efficiency of the spectral chain. This loss in efficiency is in addition to the effects of the paralyzable deadtime.

\begin{figure}
    \centering
    \includegraphics{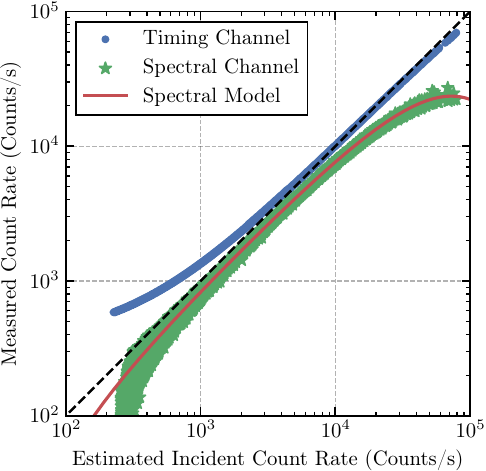}
    \caption{Measured count rates from timing chain (circles) and spectral chain (asterisks) plotted against estimated incident count rate derived from deadtime-corrected timing chain data.
    The spectral chain exhibits both paralyzable deadtime effects (\SI{13.65}{\micro\second}) and reduced detection efficiency (\SI{88.83}{\percent}), with the red solid curve representing the fitted model.
    }
    \label{fig:onboard_deadtime}
\end{figure}

Analysis of on board data, comparing spectral and timing count rates, supports this model (Figure~\ref{fig:onboard_deadtime}). Fitted parameters indicate a spectral chain efficiency of \SI{88.83}{\percent}, an additional spurious count rate of about \SI{364.44}{\counts \per \second} in the timing chain, and a paralyzable deadtime of \SI{13.65}{\micro\second} for the spectral chain. The incident count rate is derived by subtracting the spurious contribution from the timing chain data and correcting for its \SI{1.6}{\micro\second} deadtime.

\begin{equation}
m_{\text{timing}} = (n + 364.44) \cdot e^{-n \cdot 1.6\,\mu\text{s}}    
\end{equation}
\begin{equation}
m_{\text{spectral}} = 0.8883 \cdot n \cdot e^{-n \cdot 13.65\,\mu\text{s}}    
\end{equation}

The spectral chain is, therefore, impacted by both its intrinsic paralyzable deadtime and a reduction in efficiency due to periodic reset pulses, while the timing chain is affected by its own deadtime and the presence of extra spurious counts. To accurately determine the true solar X-ray count rate, the timing chain data must be corrected for its paralyzable deadtime (\SI{1.6}{\micro\second}), and the estimated spurious count contribution must be subtracted. The resulting corrected timing chain rates are then used to correct the spectral chain data, estimating the true X-ray spectrum. The deadtime correction is modelled as energy-independent, assuming minimal pile-up events. 

\subsection{Cross Calibration}
During the performance verification phase, solar X-ray observations from NOAA's XRS instrument on board GOES-16 in geostationary orbit \cite{Woods2024}, and ISRO's XSM, on board Chandrayaan-2 in lunar orbit \cite{Mithun2020_xsm} were used for cross-calibration.

The GOES-XRS instrument is a full-disk solar photometer that measures flux in two broadband channels: XRS-A (\SIrange{0.5}{4}{\angstrom}; \SIrange{3.1}{24.8}{\keV}) and XRS-B (\SIrange{1}{8}{\angstrom}; \SIrange{1.5}{12.4}{\keV}). Given the lower energy threshold of SDD2 at \SI{2}{\keV}, cross-calibration is performed in the overlapping \SIrange{3.1}{24.8}{\keV} range using the XRS-A channel. After applying deadtime corrections to the SDD2 data, count rates are converted to physical flux units using the detector's effective area response.

A comparison of XRS-A and SoLEXS fluxes at a 1-minute cadence shows a linear correlation across intermediate flux levels (\SIrange{2e-6}{1e-4}{\watt\per\metre\squared}), validating the deadtime correction and absolute radiometric calibration (Figure~\ref{fig:xrsa_solexs_scatter}). 
However, at lower flux levels, SDD2 measurements are systematically \SI{\approx 15}{\percent} lower, while during flare peaks they are higher. This trend is consistent with the known limitations in GOES-XRS's flat-spectrum assumption for flux derivation (see, for example, \cite[Section~5.3]{Schwab2020} and \cite[Figure~6]{Ng2024}).
Figure~\ref{fig:xrsa_solexs_lc} shows an example light curve that demonstrates the temporal alignment and these systematic flux deviations between the two instruments.

\begin{minipage}[t]{0.46\textwidth}
    \vspace{0.5cm}  
    \begin{figure}[H]
    \centering
    \includegraphics[width=3.15in]{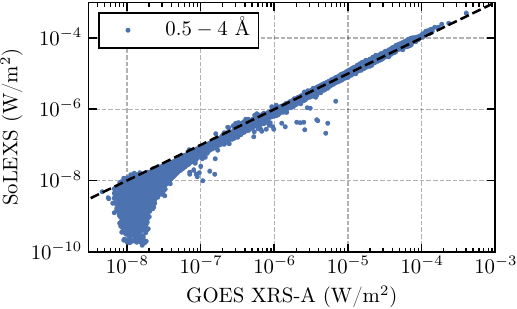}
\caption{Scatter plot of GOES-XRS-A flux versus SoLEXS SDD2 flux at 1-minute cadence during April-May 2024. }
    \label{fig:xrsa_solexs_scatter}
\end{figure}
\end{minipage}
\hspace{0.5cm}
\begin{minipage}[t]{0.46\textwidth}
    \vspace{0cm}  
    \begin{figure}[H]
    \centering
    \includegraphics[width=3.05in]{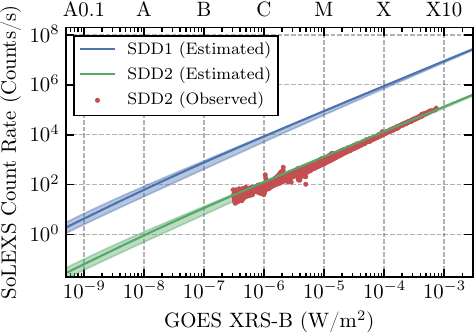}
    \caption{GOES class versus SDD2 count rate (red) and simulated count rates for SDD1 and SDD2 (blue and green) from isothermal plasma models.}
    \label{fig:xrsb_solexs_scatter}
\end{figure}
\end{minipage}

\begin{figure}[H]
    \centering
    \includegraphics[width=0.7\textwidth]{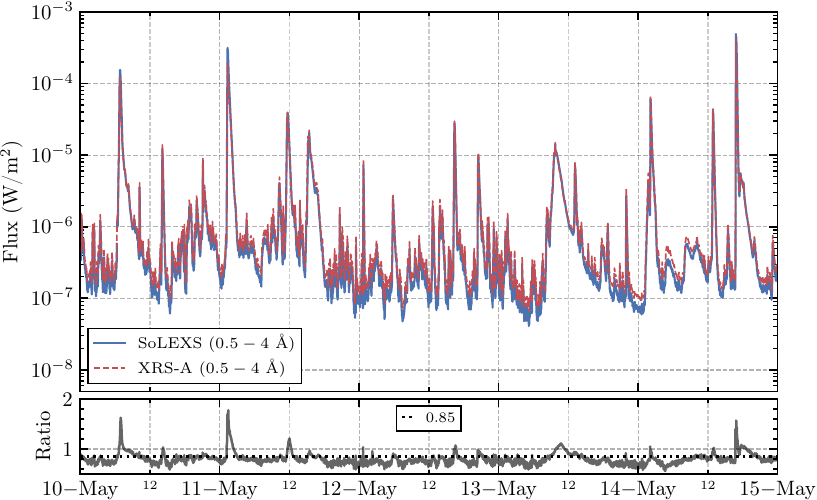}
    \caption{Light curves observed by GOES-XRS-A (dashed red) and SoLEXS SDD2 (blue) during 10-14 May 2024.
    The bottom panel displays the SoLEXS SDD2 to XRS-A flux ratio, illustrating the systematic deviations: approximately \SI{15}{\percent} lower during quiet periods and higher flux during flare peaks.
    }
    \label{fig:xrsa_solexs_lc}
\end{figure}

XRS-B flux, which is used for X-ray flare classification, is compared with the observed and estimated SDD2 count rate, as shown in Figure~\ref{fig:xrsb_solexs_scatter}. The count rates were estimated by simulated isothermal model (CHIANTI database) spectrum and incorporating the empirical temperature-XRS-B flux relationship from Ref.~\citenum{Battaglia2005}.

The XSM instrument on Chandrayaan-2 is a full-disk spectrometer operating in the \SIrange{1}{15}{\keV} range, using a \SI{450}{\micro\metre} SDD with specifications similar to SoLEXS.
While the detectors and spectral redistribution characteristics are comparable, the main differences are: XSM's aperture is \SI{0.367}{\milli\metre\squared}, approximately \num{3.46} times larger than that of SDD2 (\SI{0.1065}{\milli\metre\squared}) and it uses a standard \SI{8}{\micro\metre} Beryllium entrance window, whereas SoLEXS uses an \SI{8}{\micro\metre} DuraBeryllium Plus window.
Observations during the April 2024 Dawn-Dusk season, when lunar orbital geometry ensured full solar visibility and minimal off-axis illumination errors (\SI{<1}{\percent})\cite{Mithun2020}, were used to radiometrically cross-calibrate SoLEXS with XSM over several broadband energy ranges.

\begin{figure}
    \begin{minipage}[t]{0.48\textwidth}
        \centering
    \includegraphics{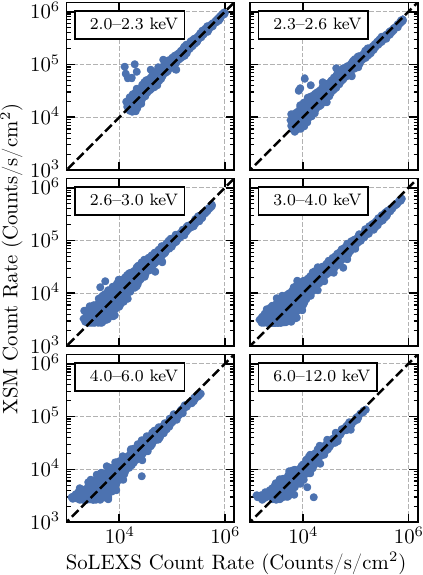}
    \caption{Scatter plot of XSM normalized count rate versus SoLEXS SDD2 normalized count rate across six different energy bands showing linear correlation.}
    \label{fig:xsm_solexs_scatter}
    \end{minipage}
    \hfill 
    \begin{minipage}[t]{0.49\textwidth}
        \centering
    \includegraphics{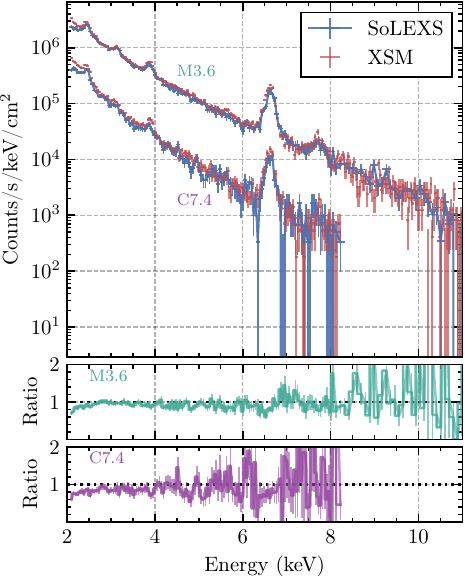}
    \caption{Spectral comparison during M3.6-class flare (SOL2024-04-23T03:19) and C7.4 flare (SOL2024-04-23T21:11).}
    \label{fig:xsm_solexs_m36}
    \end{minipage}
\end{figure}

Assuming minimal spectral redistribution differences between the two SDD-based systems, the deadtime corrected count rates divided by their respective energy-dependent, theoretical effective area (ARF) was compared (Figure~\ref{fig:xsm_solexs_scatter}). 
the count rate ratio dips to \num{\approx 0.9} in the \SIrange{2}{2.3}{\keV} energy range while remaining closer to unity at higher energies. This discrepancy at low energies, which persists even when the lower transmission of Dura Beryllium Plus is included in SoLEXS ARF, points to residual uncertainties in the ARF models. 
Furthermore, the ratio at lower count-rates (\SI{< 5e4}{\counts\per\second \per \centi \metre\squared}) is \num{\approx 0.95} across all energies, while at higher count rates, it is closer to unity. This suggest residual uncertainties in deadtime correction algorithm. 
This trend is confirmed by direct spectral comparisons. Figure~\ref{fig:xsm_solexs_m36} provide illustrative comparisons of spectra during the peak phases of two representative flares: M3.6 (SOL2024-04-23T03:19) and C7.4 (SOL2024-04-23T21:11). 
The top panels display XSM (\SI{33}{\electronvolt\per\channel}) and SoLEXS (\SI{47.75}{\electronvolt\per\channel}) spectra, normalised to respective energy-dependent effective areas. The bottom panels quantifies the XSM to SoLEXS spectral ratio.
For the intense M3.6-class flare, the ratio is close to unity (overall ratio of \num{1.06}) above the low-energy dip, consistent with the high count rate result from the scatter plot. For the weaker C7.4-class flare, the overall ratio is \num{0.91}, consistent with the low count rate result. 
Minor deviations are also present near the Fe-line complex due to slight differences in spectral resolution and at higher energies due to Poisson noise.
This detailed cross-calibration validates the SoLEXS response model to within \SI{10}{\percent} and quantifies the residual uncertainties in the ARF and deadtime corrections.

\subsection{Spectral Fitting}
The most important radiation mechanisms producing X-ray emission in SoLEXS's energy range are thermal bremsstrahlung continuum and line emission from highly ionized atoms of various elements in the solar corona. SoLEXS data can be used for the diagnosis of physical parameters, such as plasma temperature, emission measure, and elemental abundances.

\begin{figure}
    \centering
    \includegraphics{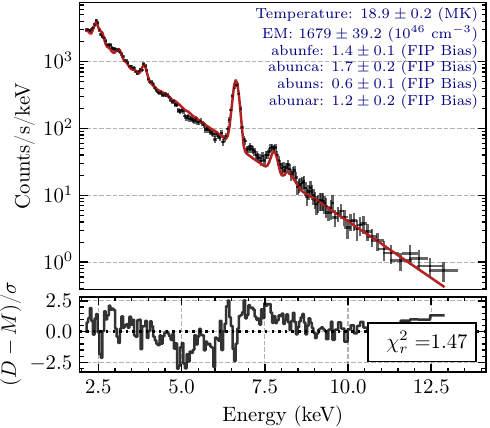}
    \caption{Observed SoLEXS spectrum (black points with error bars) fitted with an isothermal model \texttt{f\_vth\_abun} (red curve). The spectrum is integrated during peak of M6.7 class flare (SOL2024-02-12T03:46:45 to 03:47:15). Best-fit parameters
    are indicated in the annotation. The lower panel displays the residuals, with a reduced $\chi^2$ value of \num{1.47}. }
    \label{fig:fit_spec}
\end{figure}

As mentioned in Section~\ref{section:instrument_response}, the measured spectrum $P(c)$ is a convolution of the incident source spectrum $F(E)$ with the instrument response functions, as shown in Equation~\ref{eq:response}.
Determining the source photon spectrum $F(E)$ from the observed spectrum $P(c)$ and instrument responses $R(c, E)$ and $A(E)$ involves inverting this equation. However, inversion is generally not feasible and tends to yield non-unique and unstable results due to minor fluctuations in $P(c)$ \cite{xspec_guide}. An alternative approach is forward fitting, where the model spectrum $F(E)$, expressed in terms of a few parameters, is convolved with the instrumental response and fitted with the observed spectrum to get the desired fit statistics. The resulting parameters are considered ``best-fit'' parameters, which best represent the source spectrum.
The most commonly utilized fit statistic for determining the best-fit model is $\chi^2$, which is defined as follows:

\begin{equation}
 \chi^2=\sum \frac{\left(P(c)-P_{model}(c)\right)^2}{(\sigma(c))^2},   
\end{equation}

where $\sigma(c)$ is error in for photon counts in channel $c$. 
Solar flare spectra in soft X-ray range are modelled using isothermal emission models, where parameters such as temperature, emission measure, and elemental abundances are considered. The isothermal model spectrum is computed using the sunkit-spex (\url{https://github.com/sunpy/sunkit-spex}) python package. The \texttt{f\_vth\_abun} model defined in the package uses thermal bremsstrahlung for continuum and CHIANTI atomic database for emission line calculation. The sherpa fitting application \cite{sherpa_416, sherpa_paper} is used to fit the model spectrally to the observed spectrum to get the best-fit parameters. Figure~\ref{fig:fit_spec} shows a fitting example of the SoLEXS spectrum with an isothermal model with temperature, emission measure (EM), and elemental abundances of Iron, Calcium, Argon, and Sulphur.

\section{Summary}
The Solar Low-Energy X-ray Spectrometer (SoLEXS) on board Aditya-L1, is specifically designed to provide contiguous and uninterrupted solar X-ray observations from the Sun-Earth L1 Lagrange point
in \SIrange{2}{22}{\keV} energy range. With it's temporal resolution of \SI{1}{\second} and an energy resolution of $\sim$ \SI{170}{\electronvolt} at \SI{5.9}{\keV}, it is capable of addressing key science objectives of characterising coronal plasma parameters, understanding the solar coronal heating mechanisms and flare energetics through soft X-ray spectroscopy.

It consists of a pair of Silicon Drift Detectors (SDDs) with distinct aperture areas that can handle the vast dynamic range of solar X-ray intensities,
from powerful X-class events (as demonstrated in-flight by SDD2) to faint A-class microflares (the design goal for the high-sensitivity SDD1).

Ground calibration procedures were performed to characterise the instrument's spectral response, energy-channel relationship, energy resolution, deadtime effects, and collimator response. 
Thermo-vacuum testing further validated the stability of calibration parameters across the operational temperature range and functionality of the the instrument in space environment.

SoLEXS was powered on in October 2023 and successfully deployed its aperture mechanism in December 2023. Subsequent analysis of on board data facilitated 
the on board calibrations along with cross-calibration with GOES-XRS and XSM which confirmed SDD2's radiometric calibration and demonstrated consistent spectral profiles, thereby validating its capability for accurate plasma diagnostics. SoLEXS data can be effectively analysed using isothermal emission models to derive fundamental plasma parameters such as temperature, emission measure, and elemental abundances.

\subsection*{Disclosures}
The authors declare that there are no financial interests, commercial affiliations, or other potential conflicts of interest that could have influenced the objectivity of this research or the writing of this paper.

\subsection*{Code and Data}
The data that support the findings of this article are available under specific access conditions. On board performance verification (PV) phase data (from January to June 2024) are not publicly released but can be provided upon request to the corresponding author. Data acquired after the PV phase, from July 2024 onwards, are publicly accessible through the Pradan portal (\url{https://pradan1.issdc.gov.in/al1}). 
The \texttt{SoLEXS\_Tools} package, which contains all ncessary calibration database and analysis software, is also available for download from this portal. Access to the portal requires user registration. 
Ground calibration datasets are similarly available upon request.

\subsection* {Acknowledgments}

Aditya-L1 is an observatory class mission fully funded and operated by the Indian Space Research Organization. The mission was conceived and realized through collaborative efforts involving various ISRO Centres and national institutes. SoLEXS was designed and developed at the Space Astronomy Group of U R Rao Satellite Centre (URSC), Indian Space Research Organization, with contributions from multiple entities within URSC. 

We extend our gratitude to the Thermal Systems Group at URSC for thermal design, analysis and implementation support, the Mechanism Group for developing the aperture door mechanism, and the Systems Integration Group for structural analysis and vibration testing. The Reliability and Quality Assurance Group ensured component qualification and up-screening, while the Assembly and Integration Team integrated the payload to the satellite. The Mission and Flight Dynamics Group provided L0 data generation support including housekeeping and SPICE data. The Indian Space Science Data Centre (ISSDC) and ISRO Satellite Tracking Centre (ISTRAC) serve as nodal points for data downloading and communication with the Payload Operation Centre (POC), with the Aditya-L1 project team coordinating overall activities, with the POC of SoLEXS at the Space Astronomy Group of URSC generating higher level data products and posting them to ISSDC for dissemination to the scientific community.

The authors gratefully acknowledge the contributions of numerous institutions and facilities that made this work possible. We extend our appreciation to the Raja Ramanna Centre for Advanced Technology (RRCAT), Indore, for providing access to the BL-16 synchrotron beamline, which was used in characterizing the spectral redistribution function. 

We acknowledge the Environment Test Facility at URSC for providing access to their \SI{1.6}{\metre} diameter vacuum chamber, which enabled us to conduct thermo-vacuum performance testing and systematic characterization of temperature-dependent calibration parameters across operational ranges.

We acknowledge the Radiopharmaceuticals Division of Bhabha Atomic Research Centre (BARC) for manufacturing the $^{55}$Fe radioactive calibration source, which is used for continuous on board performance monitoring throughout the mission lifetime. 

We acknowledge the entire Aditya-L1 mission team for their collaborative efforts in making this solar physics mission a success.







\end{spacing}
\end{document}